\documentclass[12pt]{iopart}
\usepackage{graphicx} 
\usepackage{color}
\begin{document}
\title{Numerical simulation of the effect of pellet injection on ELMs}
\author{D Chandra, A Sen}
\address{Institute for Plasma Research, Bhat, Gandhinagar 382428, INDIA\\
Homi Bhabha National Institute, Training School Complex, Anushakti Nagar, Mumbai -400085, INDIA}
\author{A Thyagaraja}
\address{Astrophysics Group, University of Bristol, Bristol, BS8 1TL, UK}
\ead{debasis@ipr.res.in}
\vspace{10pt}

\begin{abstract}
{
We report on numerical simulation studies of the dynamical behavior of edge localized modes (ELMs) under the influence of repetitive injection of pellets. In our nonlinear 2-fluid model the ELMs are excited by introducing a particle source in the confinement region and a particle sink in the edge region. The injection of pellets is simulated by periodically raising the edge density in a pulsed manner. We find that when the edge density is raised to twice the normal edge density with a duty cycle (on time:off time) of 1:2, the ELMs are generated on an average at a faster rate and with reduced amplitudes. These changes lead to significant improvements in the plasma beta indicative of an improvement in the energy confinement due to pellet injection. Concurrently, the plasma density and temperature profiles also get significantly modified. A comparative study is made of the nature of ELM dynamics for different magnitudes of edge density enhancements. We also discuss the relative impact on ELMs from resonant magnetic perturbations (RMPs) compared to pellet injection in terms of changes in the plasma temperature, density, location of the ELMs and the nonlinear spectral transfer of energies.   
}
\end{abstract}
\maketitle
%\ioptwocol
%\vspace{-0.1in}
%\begin{figure}[h]
%\begin{center}
%\includegraphics[width=6.0cm,angle=270]{tauehrspelrmp.eps}
%\hspace{0.3in}
%\includegraphics[width=4.5cm,angle=270]{BR21vstn.eps}
%\vspace{0.05in}
%\caption{\small Effect of RMPs on time evolution of Plasma Beta} 
%\label{betvst} 
%\end{center}
%\vspace{-0.5in}
%\end{figure}
\section{Introduction}
%\vspace{-0.1in}
In high confinement plasma operation of advanced tokamaks, the occurrence of Edge
Localized Modes (ELMs) is a common phenomenon and a major concern for future
devices like ITER \cite{hender2007}. The onset of ELMs has its origin in the development of steep pressure
gradients in the edge region due to the formation of a transport barrier in the high
confinement mode (H-mode) \cite{zohm1996,connor1998,connor2008,evans2008,chapman2012}. The concomitant onset of substantial edge currents lead
to strong MHD instabilities at the edge. ELMs can significantly damage the plasma facing
components and divertor plates by their very high heat flux. ELM control and ELM
mitigation are therefore the subject of much theoretical and experimental studies at the
present time \cite{evans2008,lang2004,evans2013,baylor2015}. The linear characteristics of these modes are widely believed to be
associated with the excitation of peeling/ballooning modes \cite{connor2008,snyder2004}. Their nonlinear evolution is
however quite complex and not yet fully understood. Numerical simulations provide a
powerful tool to explore some of the complex issues related to the nonlinear behavior of ELMs \cite{thyagaraja2010,huijsmans2013,becoulet2014,orain2015,xu2016,chandra2017}.
The real challenge is in understanding the physical mechanisms underlying the suppression/mitigation of ELMs by application of external perturbations in the form of RMPs or through repetitive pellet injection as observed in several experiments \cite{lang2004,evans2013,baylor2015,evans2012}. Recently a simulation study using the CUTIE code \cite{thyagaraja2000} was successful in replicating multiple ELM cycles and also demonstrated how the application of RMPs could alter the ELM dynamics \cite{chandra2017}. CUTIE (Culham Transporter of Ions and Electrons) is a two-fluid initial-boundary value, global electromagnetic nonlinear evolution 
code. The code solves on the `mesoscale', intermediate between the device size and the ion gyroradius and takes account of classical and neoclassical transport effects. It uses a periodic cylinder model of the tokamak geometry with the toroidal curvature effects of the magnetic field lines kept to first order
in the inverse aspect ratio. Poloidal mode coupling is implemented through additional curvature terms. ELMs are simulated by introducing a particle source in the confinement region and a particle sink in the edge region. CUTIE is capable of simulating multiple edge relaxation periods and has in the past also successfully modeled the L-H transition behavior in COMPASS-D \cite{thyagaraja2010}. In this work, we report on nonlinear simulation studies using CUTIE that are aimed at exploring the dynamical behavior of ELMs under the influence of repetitive injection of pellets. We also compare and contrast the effects of pellets and RMPs \cite{chandra2017} on ELMs.\\

 The paper is organized as follows. In section \ref{pel} the effects of pellets on the dynamics of ELMs are discussed.  Section \ref{comp} provides a comparison between the mitigation of ELMs by RMP \cite{chandra2017} and pellets. Spectral analyses of the nonlinear behavior of the ELMs in the no pellet, RMP and pellet cases are presented in section \ref{spec} with a view to understand the energy cascade processes in the three cases. A brief summary and some discussion on our results are given in Sec. \ref{secsum}.\\
\begin{figure}[h]
\begin{center}
\includegraphics[width=7.9cm,angle=0]{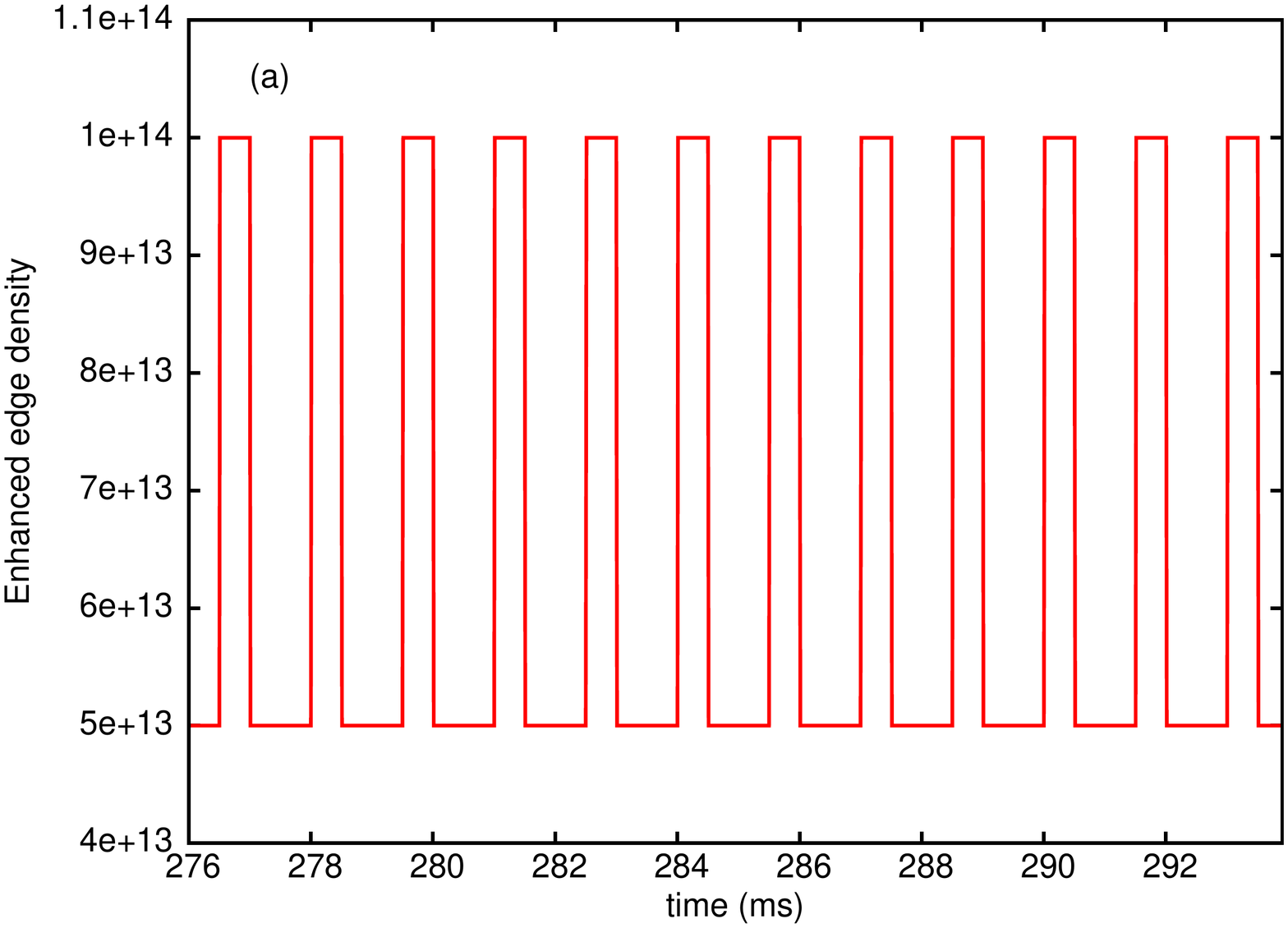}
\includegraphics[width=7.9cm,angle=0]{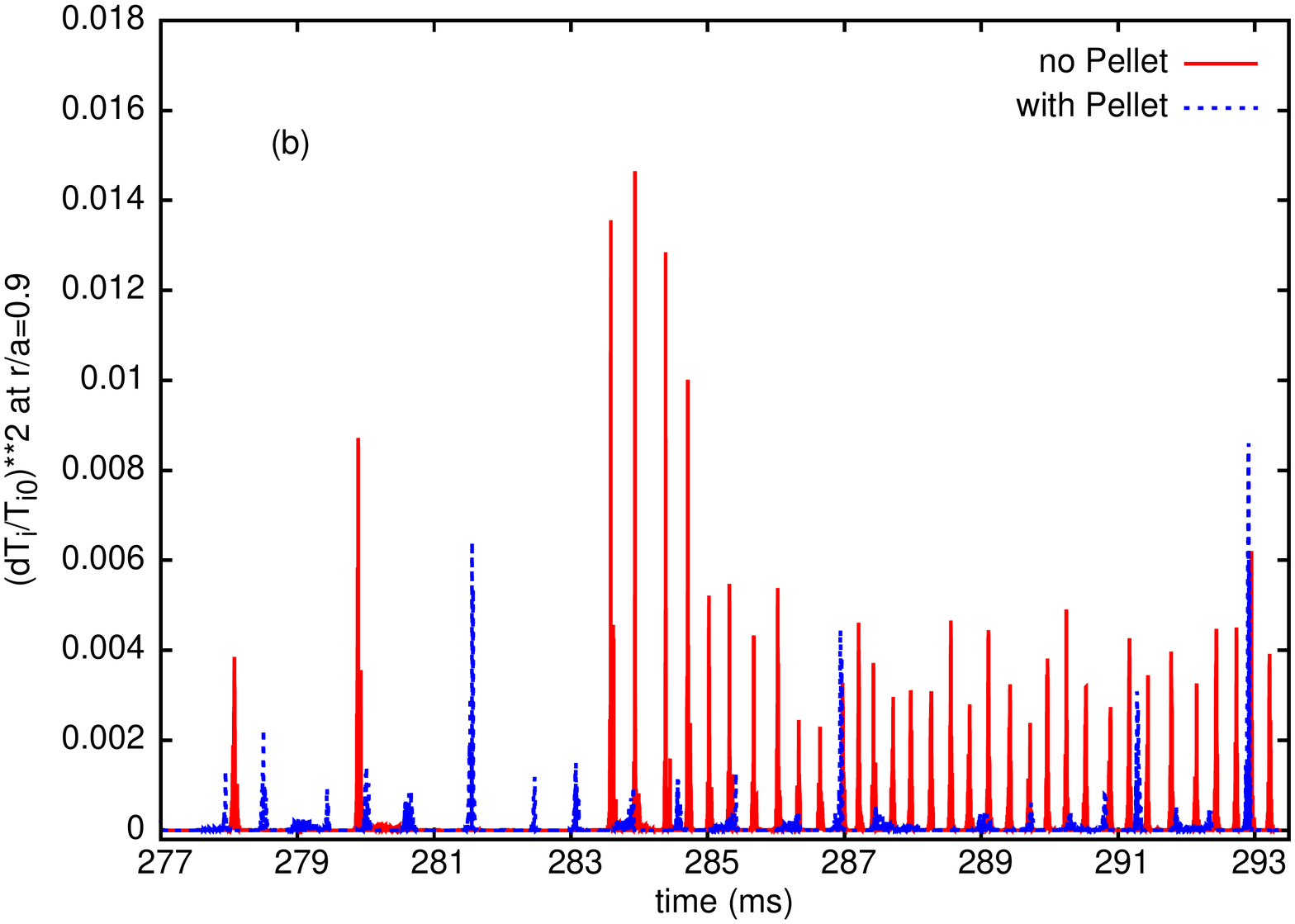}
\caption{\small (a) Time sequence of injected pellet pulses and (b) concomitant excitation of ELMs at r/a = 0.9 as shown by blue (dotted line) spikes. For comparison the excitation of natural ELMs in absence of pellet injections are shown by red (solid line) spikes.} 
\label{dtipelvst} 
\end{center}
\end{figure} 
\begin{figure}[h]
\begin{center}
\includegraphics[width=7.9cm,angle=0]{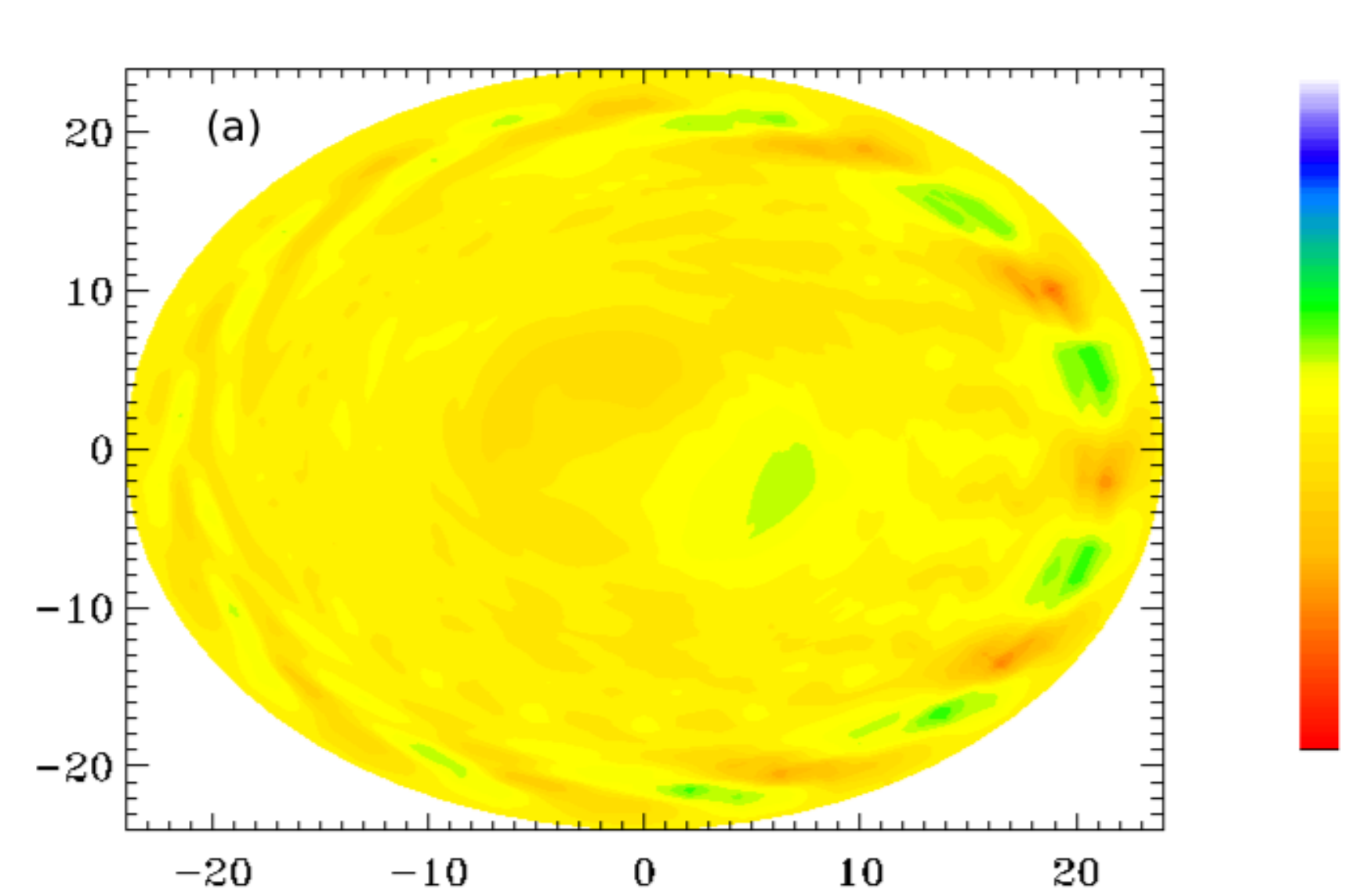}
\includegraphics[width=7.9cm,angle=0]{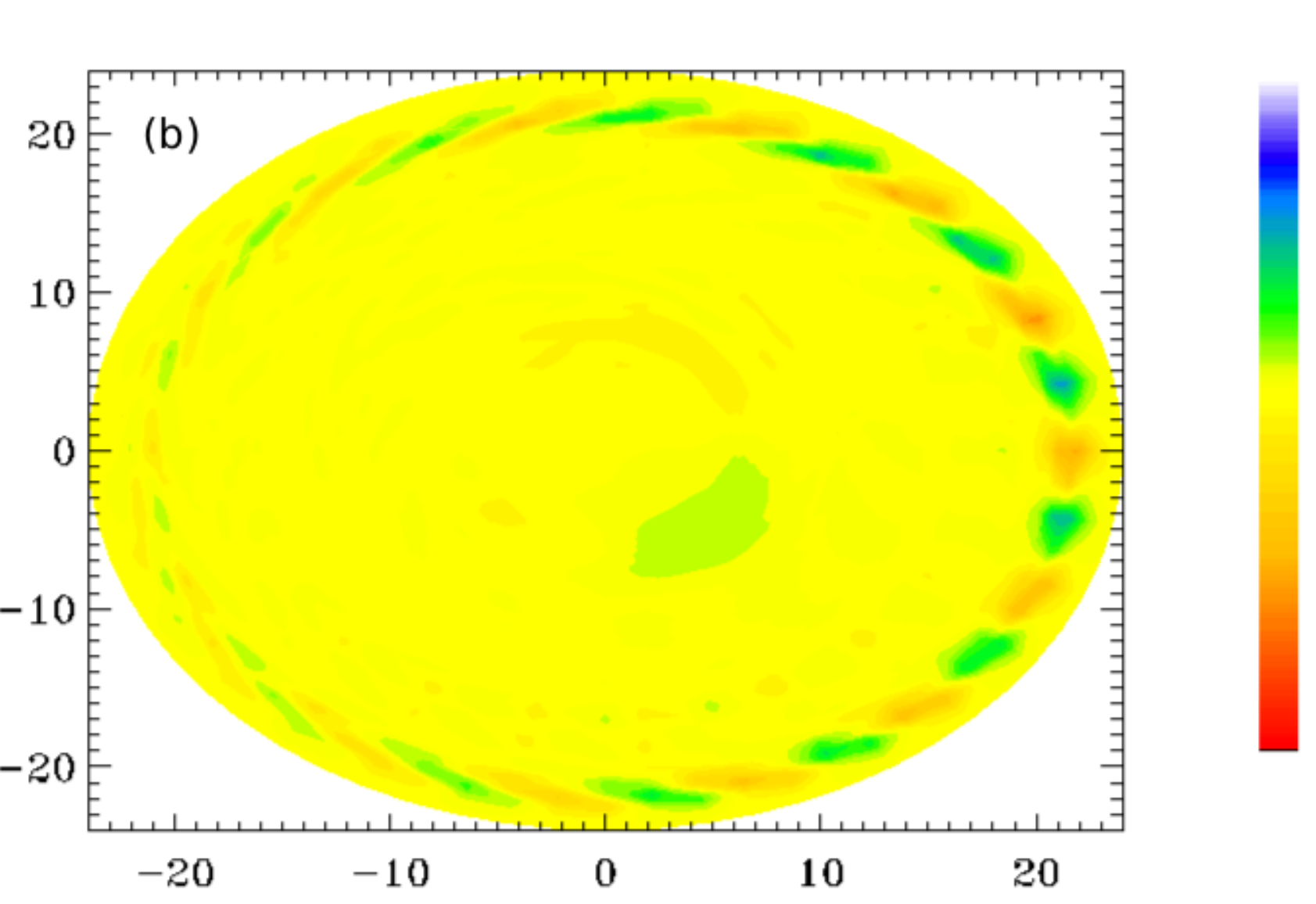}
\caption{\small Contour plots of plasma temperature during a typical ELM event (a)  with pellets and (b) without pellets.}
% Contour plots of (c) plasma density with pellets and (d) without pellets.} 
\label{contour} 
\end{center}
\end{figure} 
\begin{figure}[h]
\begin{center}
\includegraphics[width=7.9cm,angle=0]{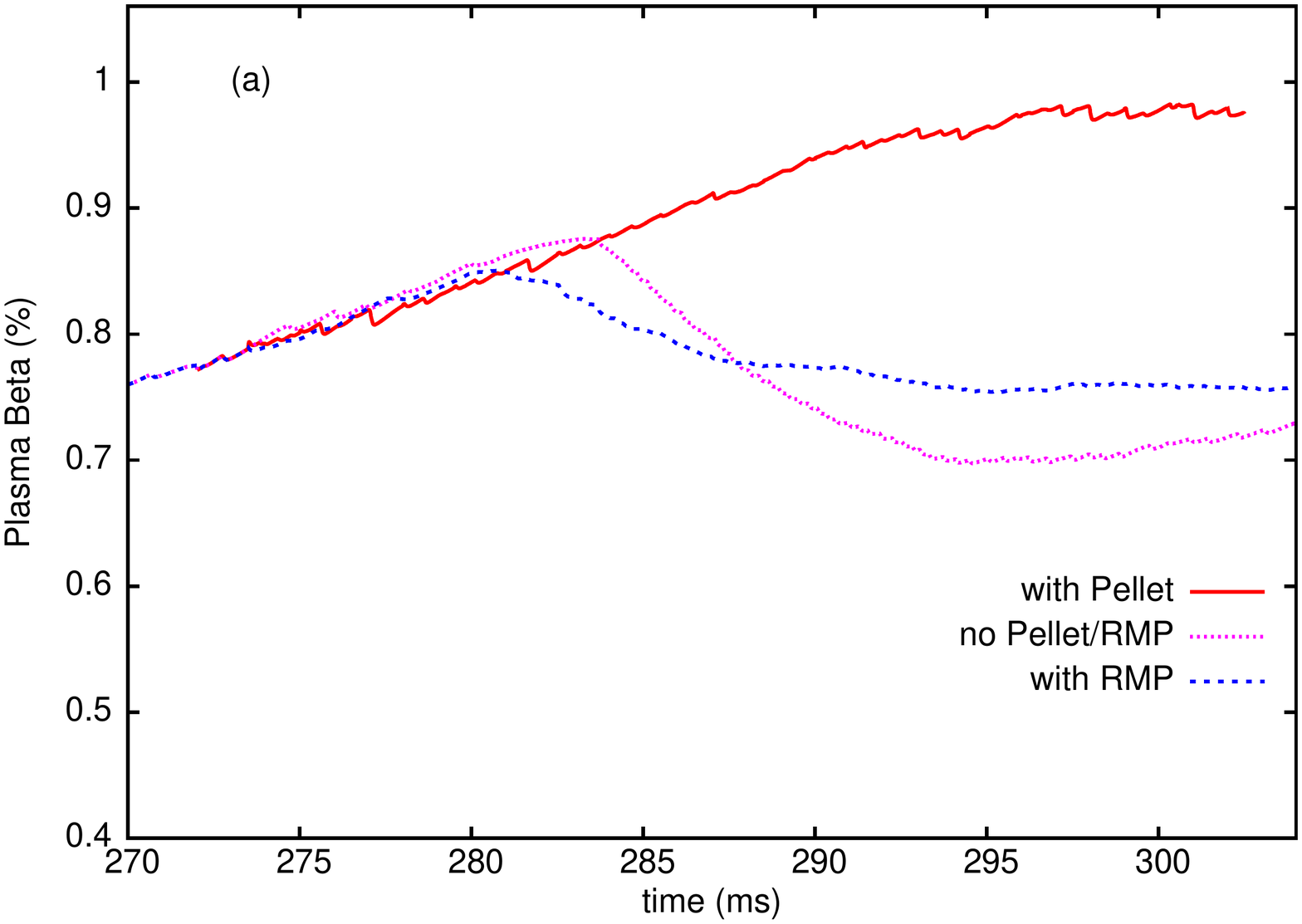}
\includegraphics[width=7.9cm,angle=0]{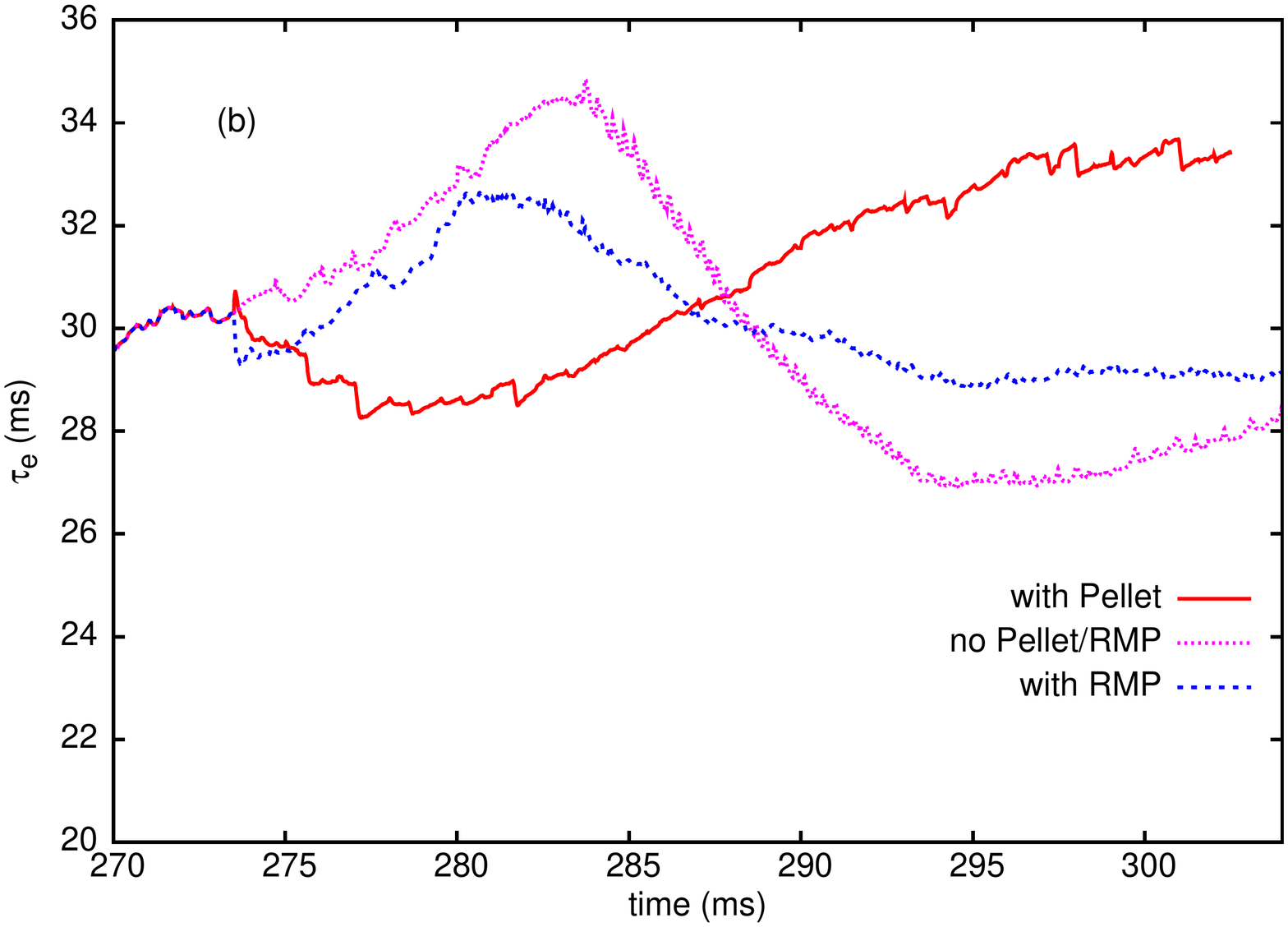}
\caption{\small Time evolution of (a) plasma beta and (b) energy confinement time in the presence of pellet injection. The red (solid) curves denote the evolution in the presence of pellets, the blue (dashed) curves denote the evolution in the presence of RMPs and the magenta (dotted) curves are for natural ELMs.} 
\label{betvstpelrmp} 
\end{center}
\end{figure} 
\begin{figure}[h]
\begin{center}
\includegraphics[width=7.9cm,angle=0]{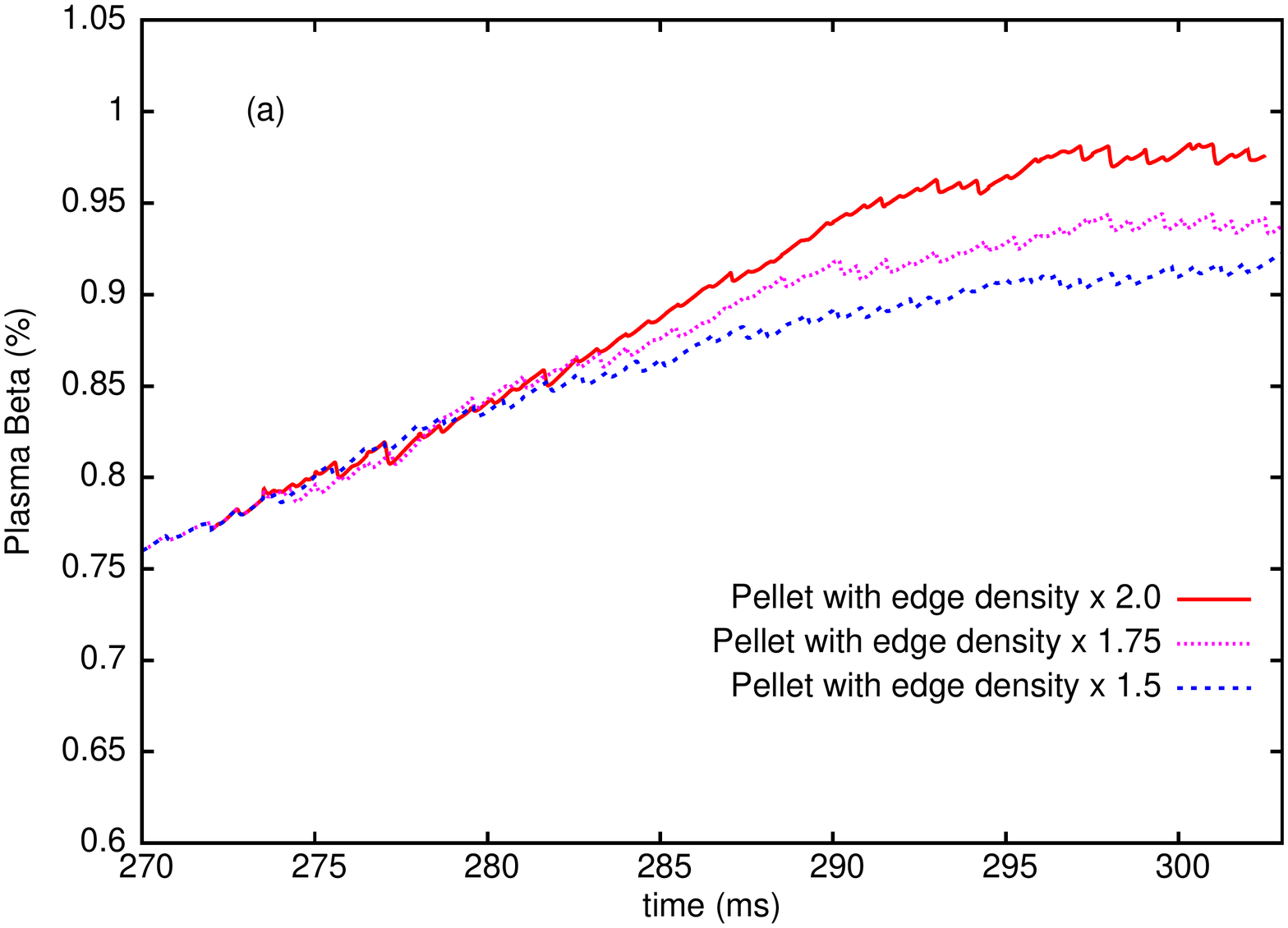}
\includegraphics[width=7.9cm,angle=0]{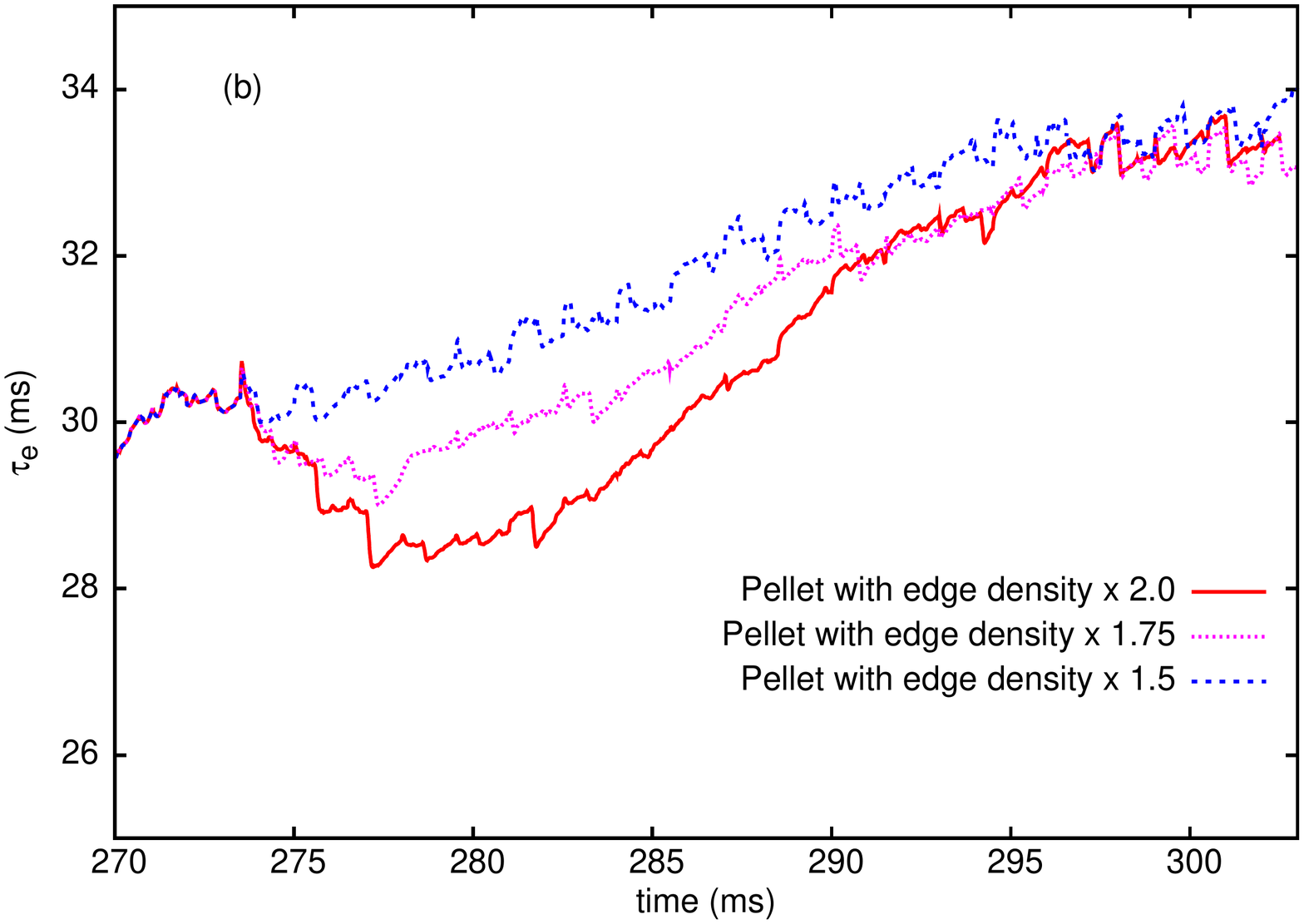}
\caption{\small Effect of enhanced edge densities due to pellet injection on the (a) plasma beta and (b) energy confinement time. The red solid curves are for enhancements that are 2 times, magenta dotted curves for 1.75 times and blue dashed curves for 1.5 times of the normal edge density.} 
\label{betvspelden} 
\end{center}
\end{figure} 
\begin{figure}[h]
\begin{center}
\includegraphics[width=7.9cm,angle=0]{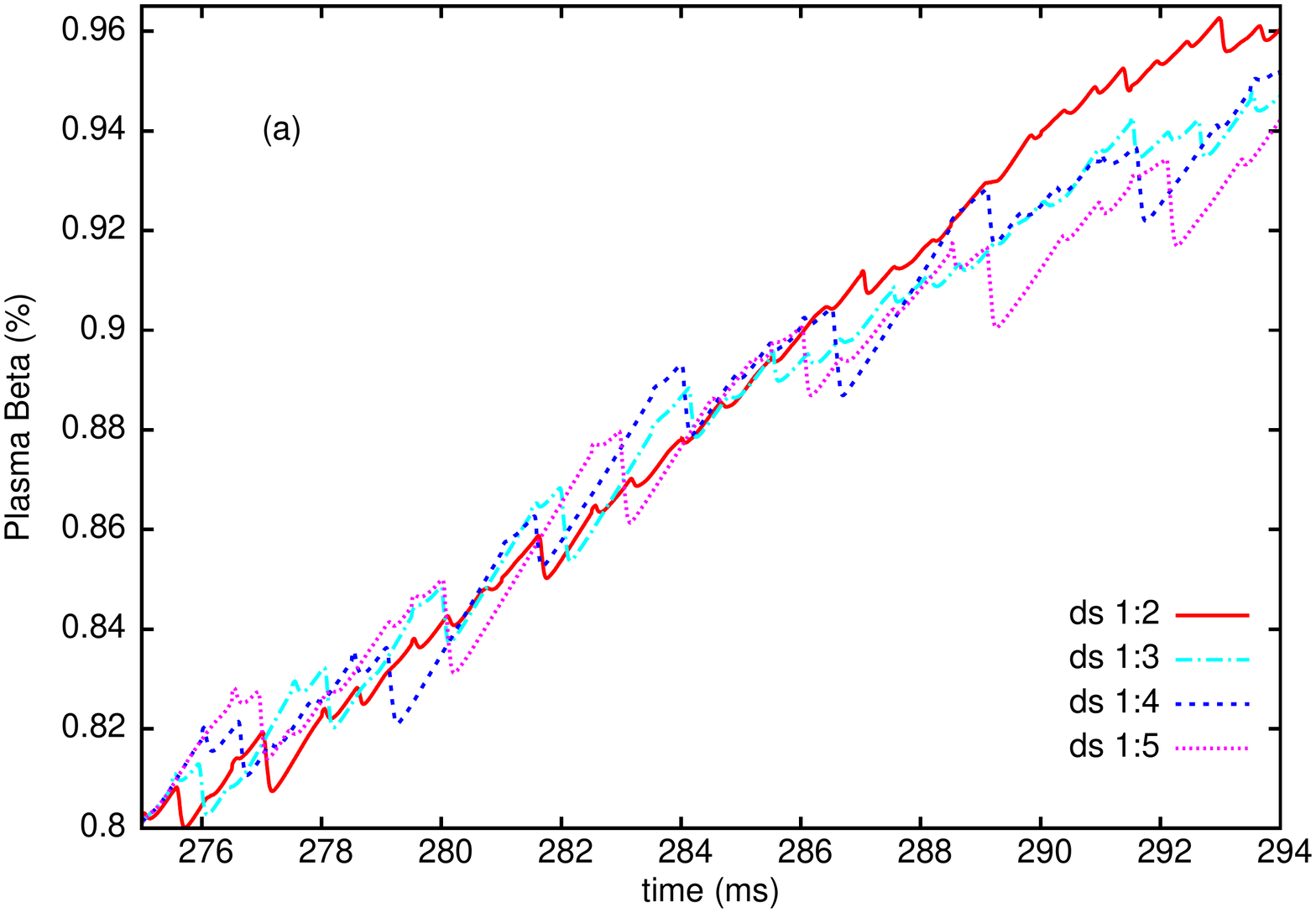}
\includegraphics[width=7.9cm,angle=0]{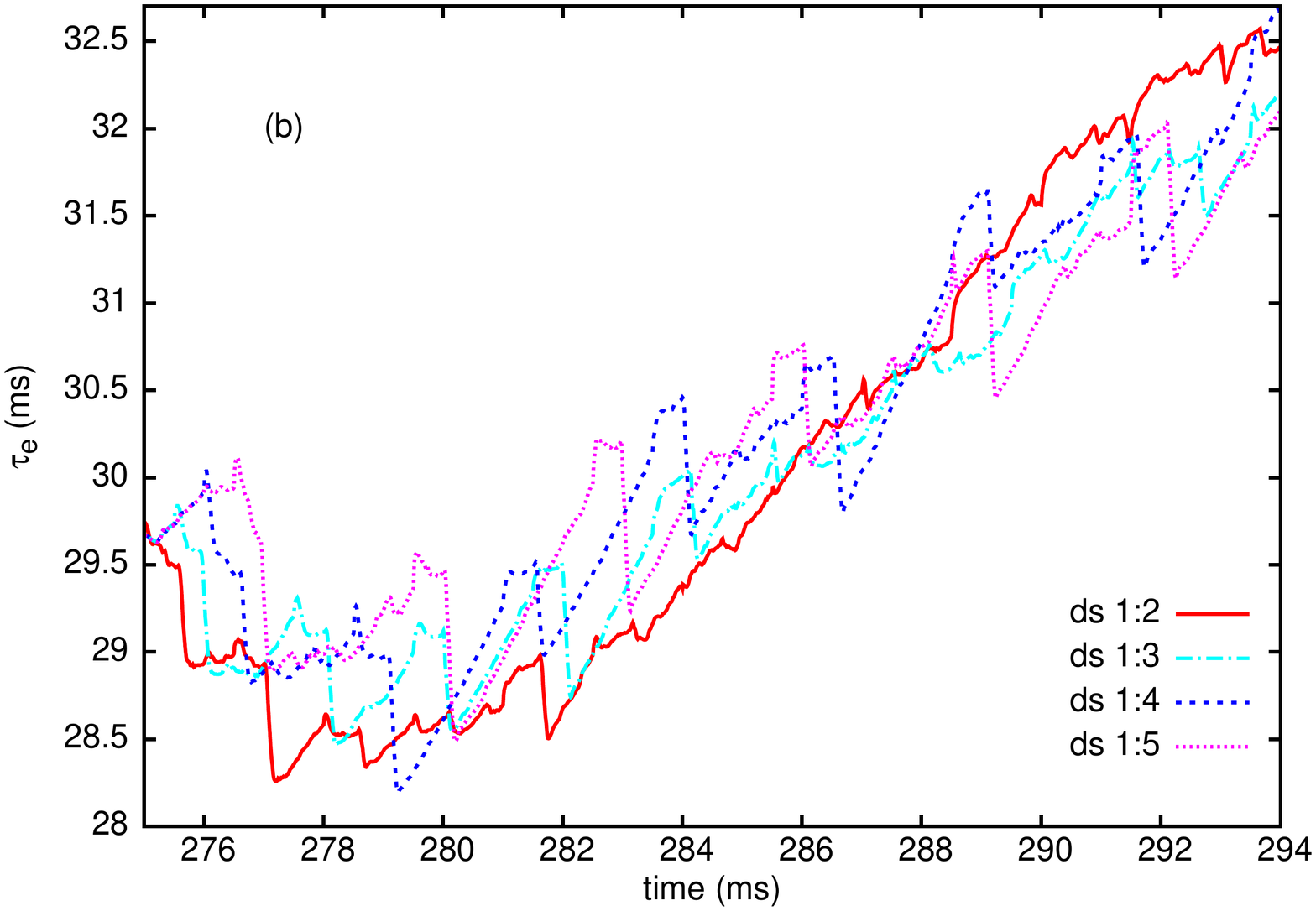}
\caption{\small Effect of variations in duty cycle of the pellet injection on the (a) plasma beta and (b) energy confinement time. The red solid curves are for duty cycle 1:2, the sky blue dot-dashed curves are for duty cycle 1:3, the blue dashed curves for duty cycle 1:4 and the magenta dotted curves are for duty cycle 1:5.} 
\label{betvspelduty} 
\end{center}
\end{figure} 
\begin{figure}[h]
\begin{center}
\includegraphics[width=7.9cm,angle=0]{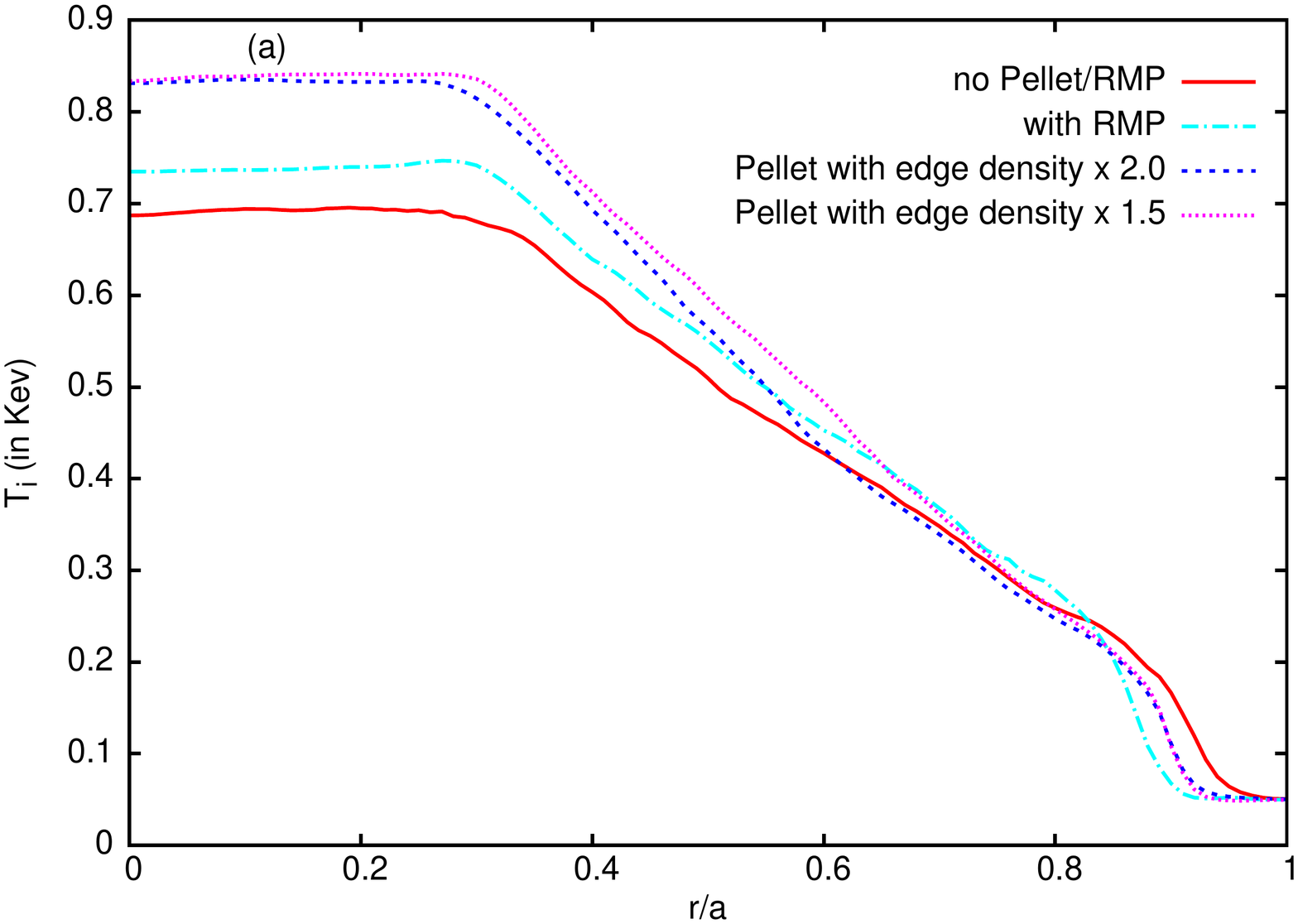}
\includegraphics[width=7.9cm,angle=0]{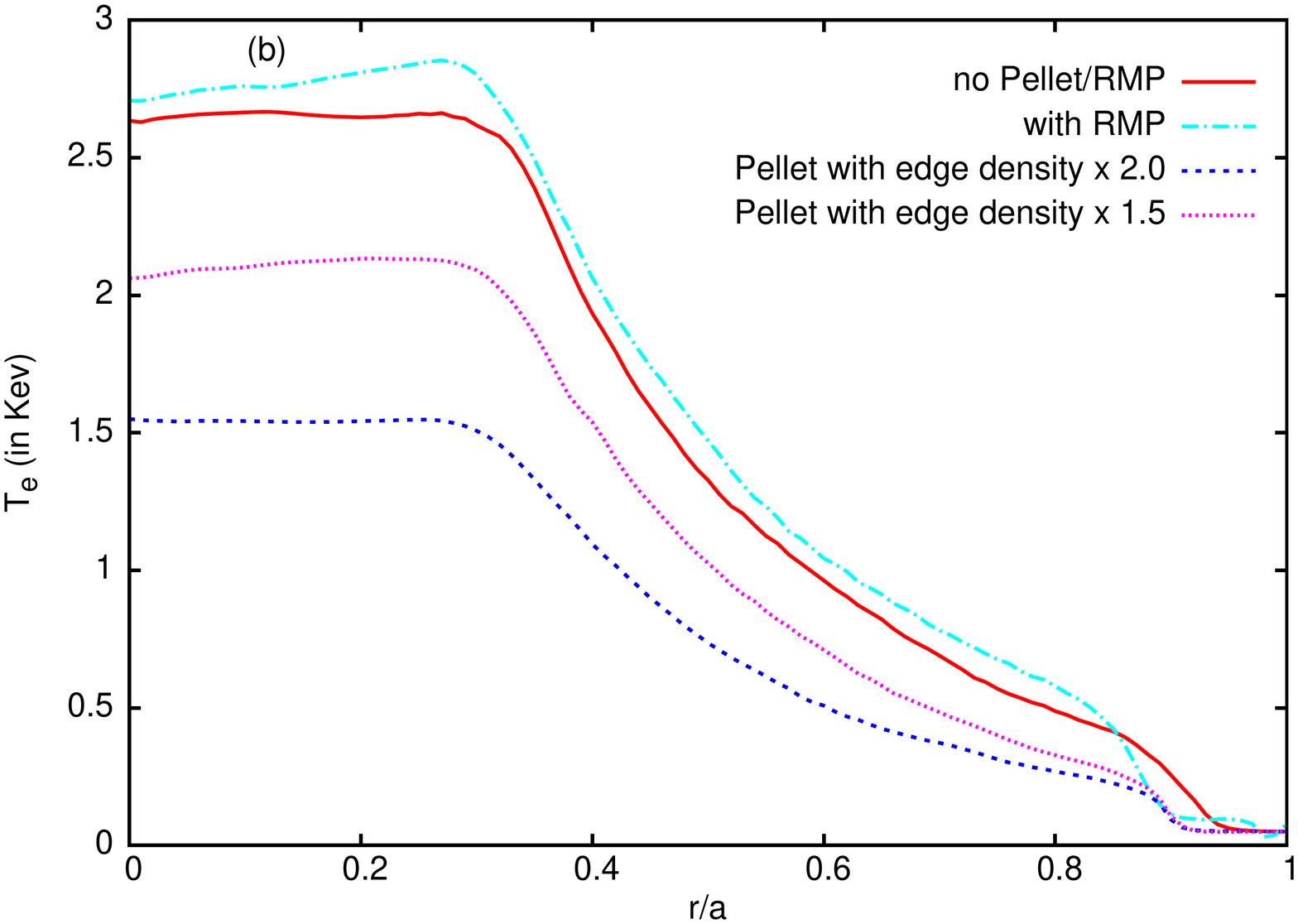}
\includegraphics[width=7.9cm,angle=0]{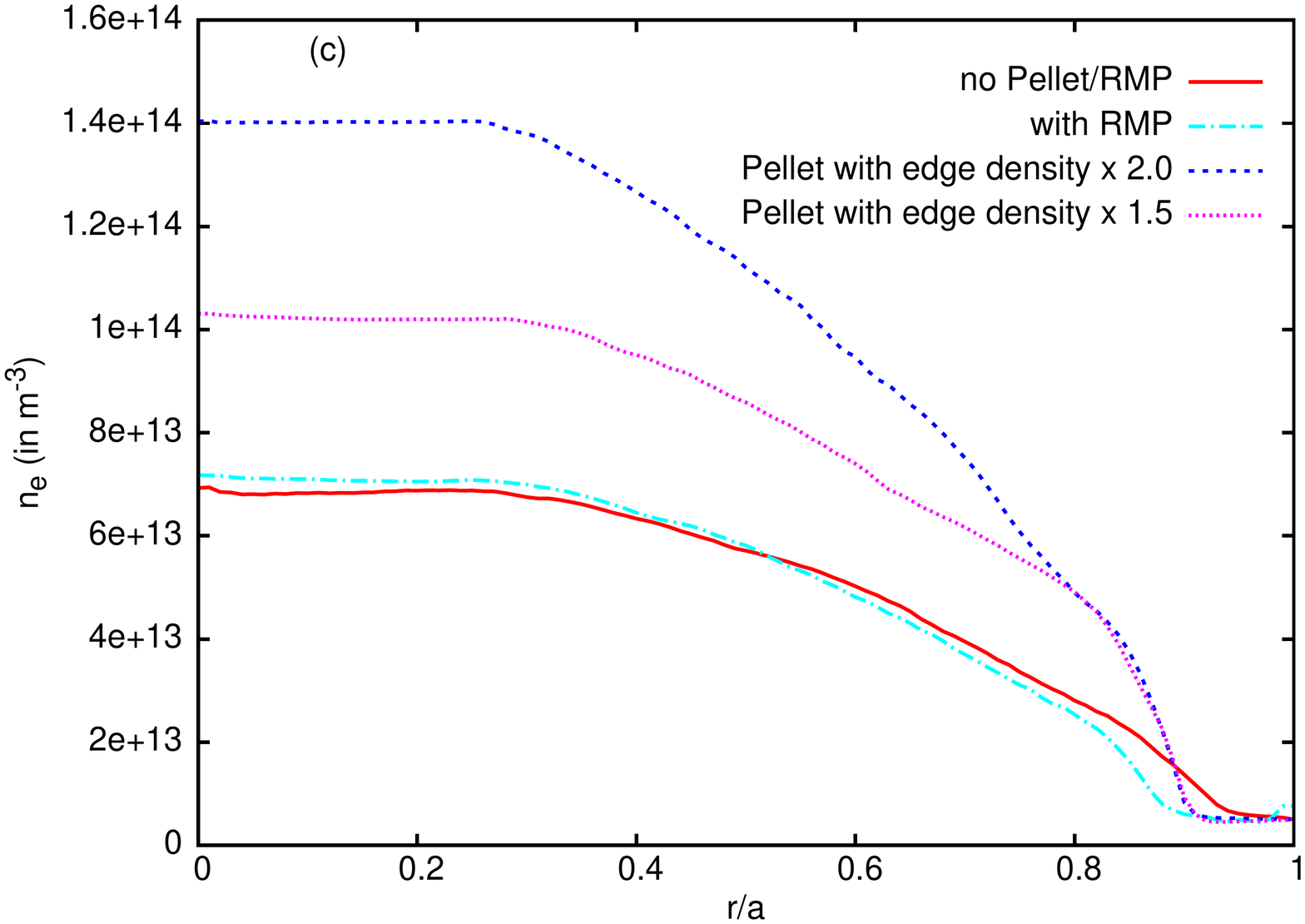}
\caption{\small Modifications of (a) ion temperature profiles, (b) electron temperature profiles and (c) electron density profiles due to different amounts of edge density enhancements (blue dashed curves for 2 times the normal edge density and magenta dotted curves for 1.5 times the normal edge density). The sky blue dot-dashed curves show modified profiles in the presence of RMPs while the red solid curves are for natural ELMs.} 
\label{profile} 
\end{center}
\end{figure}
\begin{figure}[h]
\begin{center}
\includegraphics[width=7.9cm,angle=0]{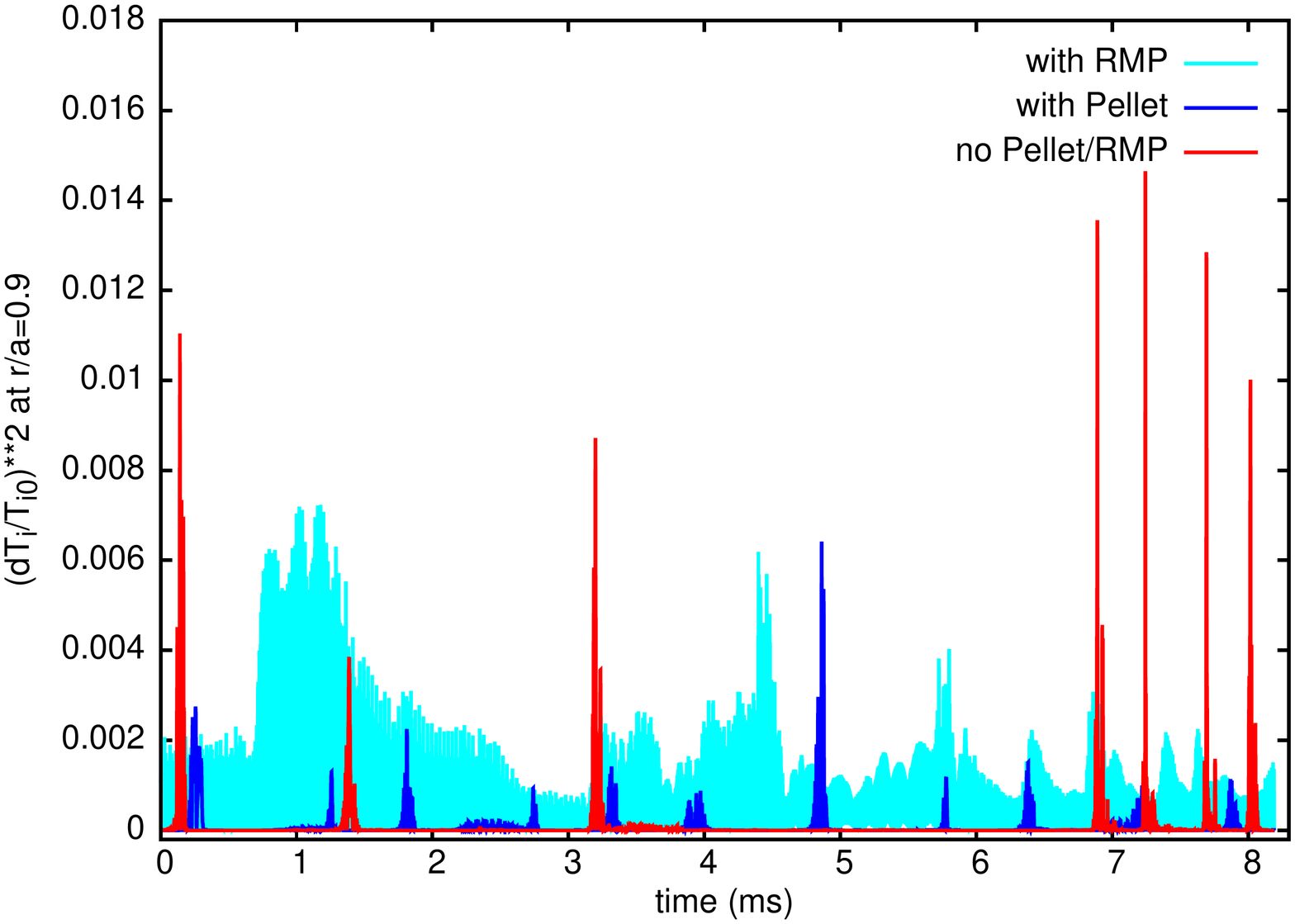}
\caption{\small The excitation of ELMs at r/a = 0.9 in the presence of pellets are shown by a blue dotted curve. For comparison, the excitation of ELMs in presence of RMPs are shown by a sky blue dash-dotted curve. The natural ELMs are shown by a red solid curve.} 
\label{dtipelrmp} 
\end{center}
\end{figure} 
%\begin{figure}[h]
%\begin{center}
%\includegraphics[width=7.9cm,angle=0]{cpltdenpelm.eps}
%\includegraphics[width=7.9cm,angle=0]{cpltdennopelm.eps}
%%\includegraphics[width=7.9cm,angle=0]{rmp.eps}
%\caption{\small Spectrum plot during ELMs with Pellets (in top left), without any Pellets/RMPs (in top right)  and with RMPs (in bottom) } 
%\label{modespec} 
%\end{center}
%\end{figure} 
%\section{Simulations of ELMs}
%\label{elms}

\section{ELM dynamics with Pellets}
\label{pel}
The injection of pellets in CUTIE is simulated by the periodic raising of the edge density in a pulsed manner. In other words the pellet is represented by a local, time-dependent particle source term to which the $n_{e0}(r, t)$ profile responds. The spatio-temporal area of the pulse models the pellet size and the repetition rate is quantified in terms of a duty cycle defined as the ratio of the `on time' and `off time' of the edge density enhancement. The `off time' physically correspond to the interval between pellet injection in an experiment. Typical results from the injection of pellets and their concomitant impact of ELM dynamics are shown in Fig.\ref{dtipelvst}. The figure shows the wave forms of the square of the edge temperature perturbations, $(dT_i/T_{0i})^2$ at $r/a=0.9$ with and without pellet injection for a machine with plasma parameters typical of COMPASS-D  with $R=0.56m$, $a=0.17m$, $B_{\rm tor}=2.07 T$, $I_{p}=242 KA$ and an electron cyclotron heating power of $P_{ECH} = 0.5 MW$. The red solid curves are for  natural ELMs (no pellet injection) and the blue dotted lines represent ELMs in the presence of pellets. The pellet size was chosen to cause an edge density enhancement by a factor of $2$ and to operate at a duty cycle of 1:2. 
As can be seen, the injection of pellets brings about a significant difference in the ELM excitations. In general there is a considerable reduction in the amplitudes of the ELMs due to the injection of pellets. There is also a qualitative change in the nature of the ELMs in the way of a large number of smaller ELMs associated with a major ELM event. The ballooning character of the ELMs is however not changed, as evident from the contour plots of the temperature perturbations in the absence and presence of pellet injections shown in Fig.\ref{contour}. 
In Fig.\ref{betvstpelrmp} we have shown the associated changes in the evolution of the plasma beta ($\beta$) and the energy confinement time. Here the energy confinment time $\tau_e = W(t)/P_{tot}$ and $\beta = W(t)/\int{B^2 dv/8\pi}$ where $W = \int{nTdv}$ is the total plasma internal energy and $P_{tot}$ is the total injected power (auxiliary plus Ohmic).
The mitigation of ELMs by pellet injection is seen to significantly improve the plasma beta similar to earlier observation in the case of application of RMP fields \cite{chandra2017}. A comparison of the two cases will be discussed in the next section. 

To further explore the efficacy of the pellet injection method of ELM mitigation, we have made simulation runs with different sizes of pellets (corresponding to different values of the edge density enhancement). We have found that with the reduction in the size of the pellet the achievable plasma beta reduces but the asymptotic energy confinement time is not significantly altered as shown in Fig.\ref{betvspelden}. 
%Further more for the limited runs made so far the duty cycle of $2:1$ seems to be the most efficient for ELM mitigation. 
We have also done simulation runs with different duty cycles of pellet injection as shown in Fig.\ref{betvspelduty}. We find that as the interval between pellets is increased there is a concomitant decrease in the
achievable beta and energy confinement time.
In Fig.\ref{profile}, we have compared the typical plasma temperature and density profiles around $t=298$ ms for runs with different pellet sizes. The central density rises substantially above the value for natural ELMs as a function of the pellet size. The central electron temperature on the other hand decreases significantly with the size of the pellet. 
%{\color{blue} This is because of the fact that the enhancement of plasma density can cause the reduction of the electron temperature by collission process \cite{liu1994}}. 
Contrary to electron temperature, the central ion temperature is seen to rise above the value for natural ELMs but is relatively independent of the size of the injected pellet. The decrease in the central electron temperature can be qualitatively understood as follows. The introduction of a pellet induces a rise in the number of electrons in the plasma which reduces the energy per electron ($T_{0e}$) since the heat source (essentially $P_{ECH}$ ) is held constant. With a rise in the electron density and a fall in the electron temperature, the electron ion thermal equilibration rate (proportional to $n_{e0} / T_{0e}^{3/2}$) necessarily rises \cite{liu1994,milora1980}. It is interesting to note that similar changes in density and temperature profiles following pellet injection have been observed in several tokamak and stellarator experiments \cite{liu1994,milora1980,bozhenkov2018}.\\

 The net result is an increase in the plasma beta as a function of increasing pellet size as seen in Fig.\ref{betvspelden}. The profiles also show the development of steep temperature and density gradients at the edge - pedestal formation - that is typical of H mode plasmas. The pedestal location also appears to shift inwards (towards the centre) for the pellet injection case as compared to the natural ELMs case. Such observations agree well with DIII-D experiments with pellet injection \cite{baylor2000}. 
 
\label{pelrmp}
\begin{figure}[h!]
\begin{center}
\includegraphics[width=7.9cm,angle=0]{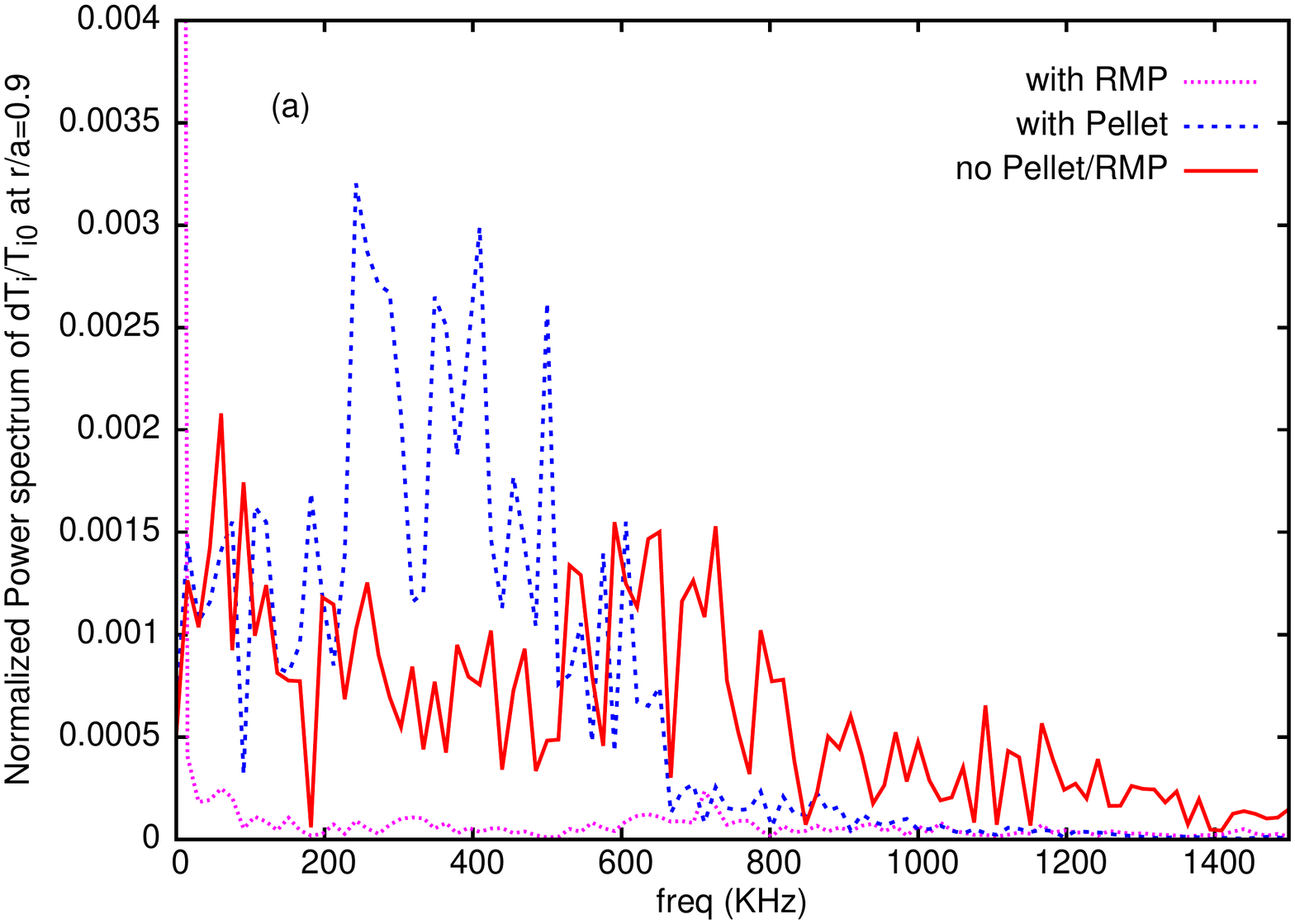}
\includegraphics[width=7.9cm,angle=0]{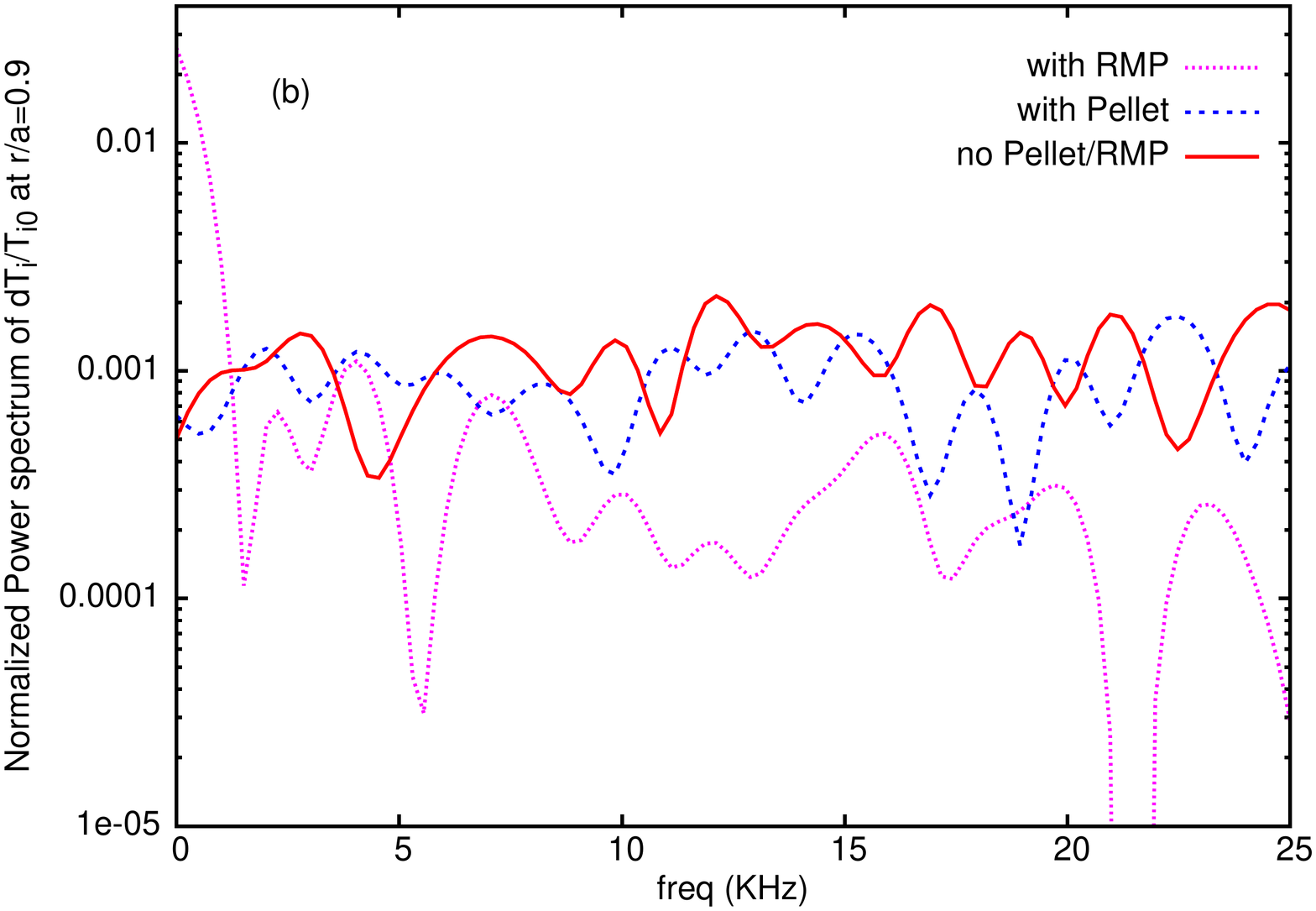}
\caption{\small Comparison of (a) Frequency power spectra of temperature fluctuations for pellet injection (blue dashed curve) and RMPs (magenta dotted curve) and natural ELMs (red solid curve). (b) Amplified plot of the low frequency portion of the power spectra.} 
\label{power} 
\end{center}
\end{figure}
\begin{figure}[h!]
\begin{center}
\includegraphics[height=6.9cm,width=7.9cm,angle=0]{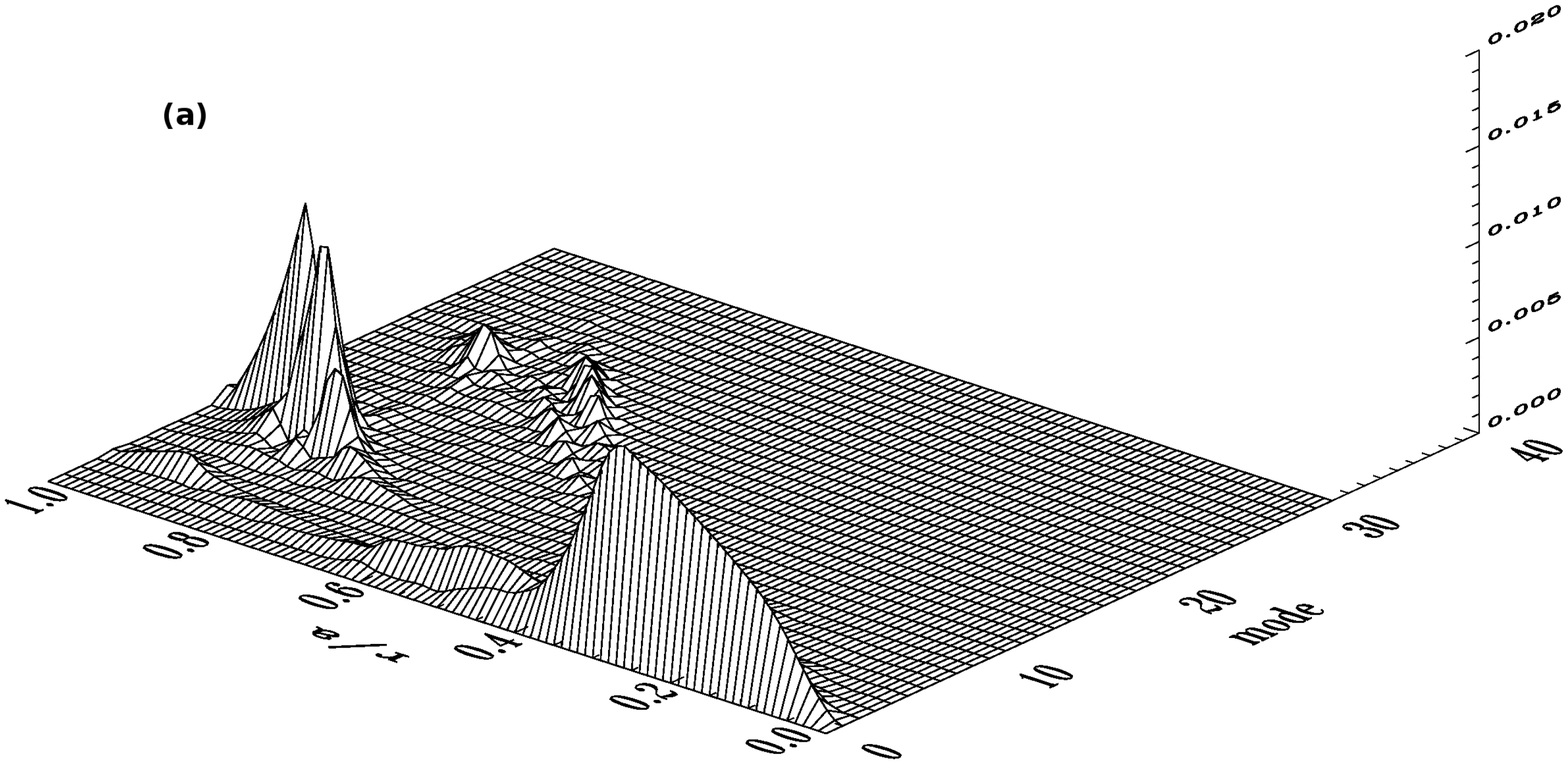}
\includegraphics[height=6.9cm,width=7.9cm,angle=0]{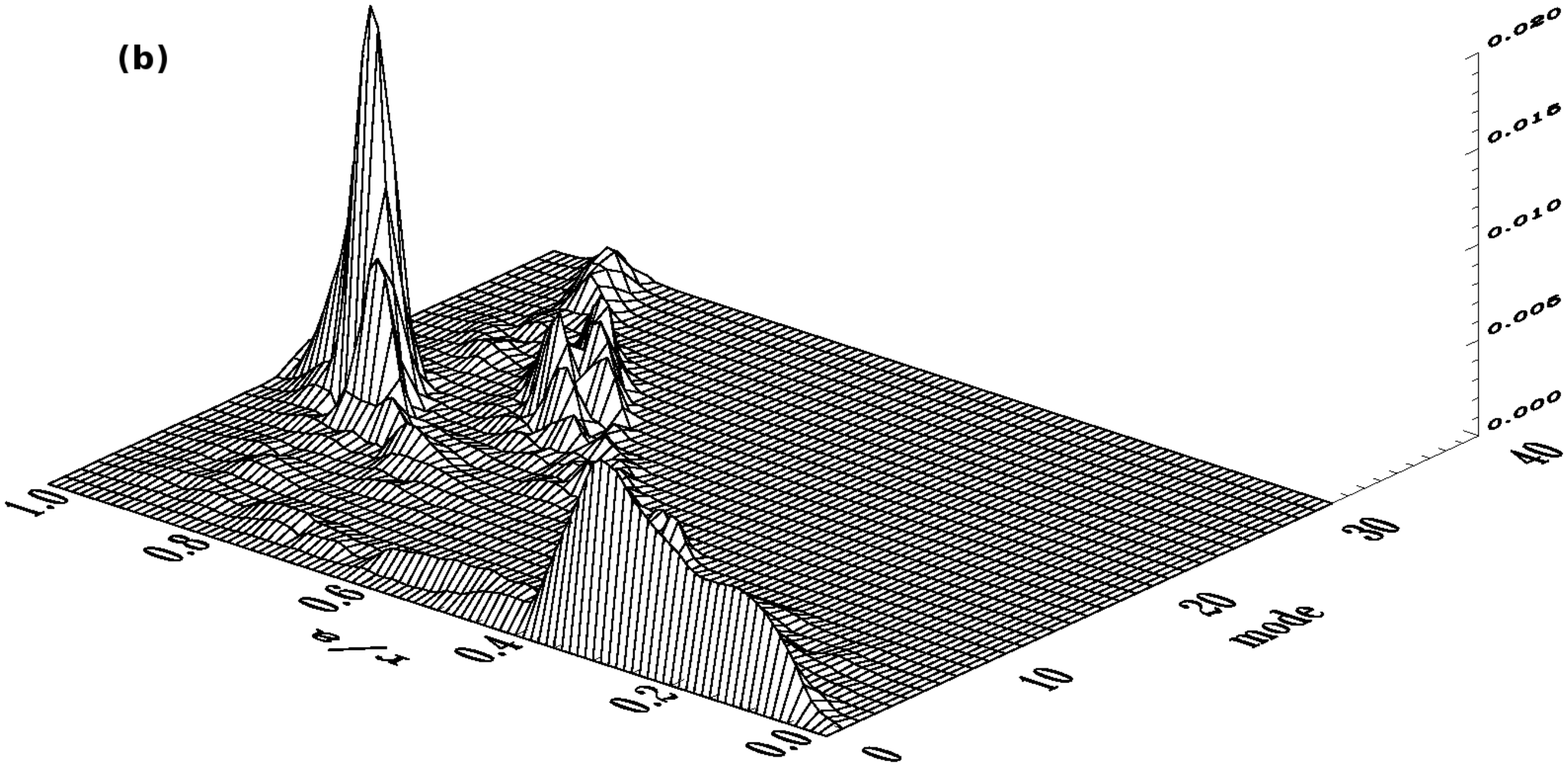}
\includegraphics[height=6.9cm,width=7.9cm,angle=0]{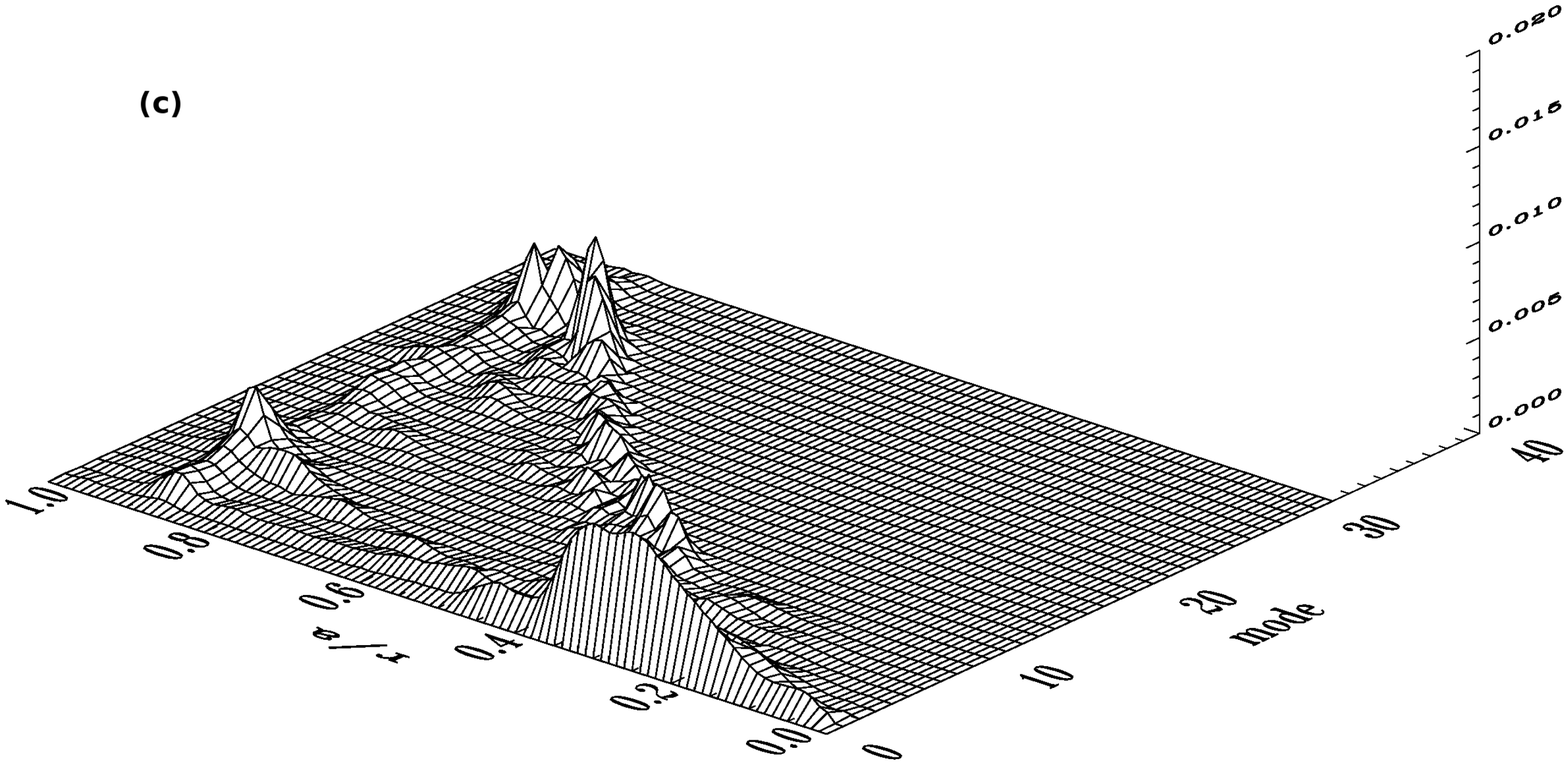}
\caption{\small Wave number spectra plots of temperature fluctuations during a typical ELM event (a) with pellets, (b) natural ELMs  and (c) with RMPs} 
\label{modespec} 
\end{center}
\end{figure} 
\section{Comparison of ELM dynamics with Pellets and RMPs}
\label{comp}
In Fig.\ref{dtipelrmp}, we have compared the wave forms of the square of the edge temperature fluctuations, $(dT_i/T_{0i})^2$,  for a pellet injection case with one where RMPs have been used. As can be seen, the two wave forms have significant differences. With the application of RMPs, the ELMs are transformed to a ``grassy'' type, whereas for the pellet case although the amplitude of the ELMs reduces significantly, they still maintain their original individual spiky character. Notable differences for the two cases are also observed in the rise of beta and energy confinement time. As seen from Fig.\ref{betvstpelrmp}, pellets lead to a higher rise of beta and better energy confinement time compared with RMPs. Pellets cause a higher rise in the central ion temperature compared to a relatively weaker rise in the case of RMPs. RMPs however cause a larger inward shift in the edge ion temperature gradient compared to pellets. 
As for the central electron temperature, RMPs bring about an increase in contrast to pellets which lead to a decrease in the central electron temperature. The central electron density increases significantly with pellets but for RMPs it registers a mild improvement over the value for natural ELMs. 
% much for RMP case comapare to natural ELMs i.e. no Pellet/RMP case. The location of edge density barrier shifting %inside for pellets compare to natural ELMs but not as much like RMPs. It looks density gradients for pellet cases %are more steeper compare to other cases.  
All these features cause a larger rise in beta for the pellet injection relative to RMPs.     
\section{Energy distribution of modes}
\label{spec}
To get some insight into the underlying nonlinear dynamics governing the behavior of ELMs, we have looked at the energy distribution of ELMs at different frequencies for natural ELMs as well as in the presence of pellets and RMPs.  The normalized power spectra for the three cases are presented in Fig.\ref{power} and they show distinct differences. In the case of unmitigated natural ELMs the power distribution is more or less the same for most of the frequency range except for a slight reduction at higher frequencies. For the pellet and RMP simulations, the high frequency components are small but they differ in the medium and low frequency ranges. In the medium range of 200-600 kHz, there is a significant amount of power for the pellet case as compared with RMPs. 
Fig.\ref{modespec} shows the energy spectrum of ELMs for different mode numbers. The nature of the spectrum for the case of pellet injection is similar to that of the no-pellet case except that it has lower amplitudes and the ELMs have relatively lower mode numbers. However their nature is different from the spectrum of RMPs. In the case of RMPs, the ELMs are pushed somewhat inside the edge and they also have significant energy for higher mode numbers.
%\vspace{-0.04in}
\section{Summary and Discussion}
\label{secsum} 
To summarize, we have carried out numerical simulations to study the impact of pellet injections on edge localized modes. The simulations have been carried out using the two-fluid CUTIE code that models ELMs by utilizing a particle source in the confinement region and a particle sink in the scrape-off layer. The injection of pellets is modeled by raising the edge density in a pulsed manner with the size of the pulse corresponding to the size of the pellet. 
%The injection rate is further quantified by a duty cycle that defines the ratio of the `on time' to the `off time' of the edge density enhancement. 
Our results show that the injection of pellets has a significant effect on the ELM dynamics by reducing the amplitude of the ELM bursts and consequently improving the plasma beta and energy confinement time of the system. Pellets also induce changes in the equilibrium profiles of the 
electron/ion temperatures and densities and modifications in the pedestal profile and location. These are further found to be sensitive to the pellet size.  A comparison with our past results on RMP induced mitigation of ELMs \cite{chandra2017} shows that pellets are more effective in raising the plasma beta and improving the energy confinement times of the system. Such an improvement of the energy confinement with the introduction of pellets has been observed in experiments as on ASDEX \cite{kaufmann1988}.

Further work is required to fully explain the results of our present work. However, certain general observations are in order: it seems clear that pellets involve a strong, time-dependent particle source which changes the density and pressure profiles. In the no-pellet and RMP simulations, the particle and energy sources are essentially fixed apart from small changes in the Ohmic heating. The RMP simulations involve changes to the poloidal magnetic field at the plasma edge due to the RMP coil currents (time-independent). This latter feature is lacking in both the reference no-pellet and the pellet simulations. We believe that RMP effects to the power spectra and to the dynamical changes to ELMs is largely due to the modulational instabilities induced by the interaction of the ELMs and the helically symmetric high-m field structure due to the RMP coils. In
the case of pellets, the strong density rises at the edge apparently leads to an inverse cascade to mid frequency ranges. This could be due to drift/two-fluid stabilisation of the MHD (ballooning-peeling modes) seen in both the no-pellet and pellet simulations. Obviously, the system is strongly nonlinear with significant interaction between the “equilibrium/axi-symmetric” profiles of density, current, electric potential and temperature and the turbulence generated by instability. The simulations reveal that the dynamics are changed depending upon  whether the RMPs or pellets are used to perturb the ELM-ing “reference”
plasma. However we have not observed a simple correlation between the ELM-ing rate with the rate of the pellet injection as observed in some experiments. These and other effects such as the influence of plasma rotation and the concurrent use of RMPs and pellets are important problems that we plan to address in the future.\\

\noindent
{\bf References}\\
\bibliographystyle{unsrt} 
\bibliography{mhd4}
\end{document}